\documentclass{PoS}
\usepackage{subfigure}
\title{Top quark event modelling and generators in the CMS experiment at the LHC}

\ShortTitle{Top quark event modelling and generators in CMS}

\author{\speaker{Bu\u{g}ra Bilin}\footnote{On behalf of the CMS Collaboration}\\
        Middle East Technical University, Ankara TURKEY\\
        E-mail: \email{bugra.bilin@cern.ch}}

\abstract{State-of-the-art theoretical predictions accurate to next-to-leading order QCD interfaced with {\sc pythia} and {\sc herwig} are tested by comparing the unfolded $t\bar{t}$  differential data collected with the CMS detector at 8 TeV and 13 TeV. These predictions are also compared with the measurements of underlying event activity distributions accompanying ${\rm t\bar{t}}$ events. Furthermore, predictions of beyond NLO accuracy in QCD are compared with the data.}

\FullConference{XXIV International Workshop on Deep-Inelastic Scattering and Related Subjects\\
		11-15 April, 2016\\
		DESY Hamburg, Germany}

\newcommand{\deleted}[1]{}
\begin{document}

\section{Introduction}
The top quark is the heaviest known elementary particle up to date, and the only coloured particle decaying before hadronisation.
The theoretical modelling of top-quarks is crucial for various measurements at the Large Hadron Collider (LHC). For example, main systematics for the top mass measurements are due to modelling of top quark events and description of fragmentation and hadronisation of bottom quarks; which can be reduced by improvements in the Monte-Carlo (MC).
In this note several CMS [1] measurements [2-8] are summarised sharing the scope of testing the predictions of the state-of-the art MC generators available with the data, and providing means of constraining uncertainties rising from these theory predictions.
\section{Comparisons of jet multiplicity and top $p_{\rm T}$ }
\label{sec:theo}

\begin{figure}[htb!]
\centering
\subfigure[]{\includegraphics[scale=0.3]{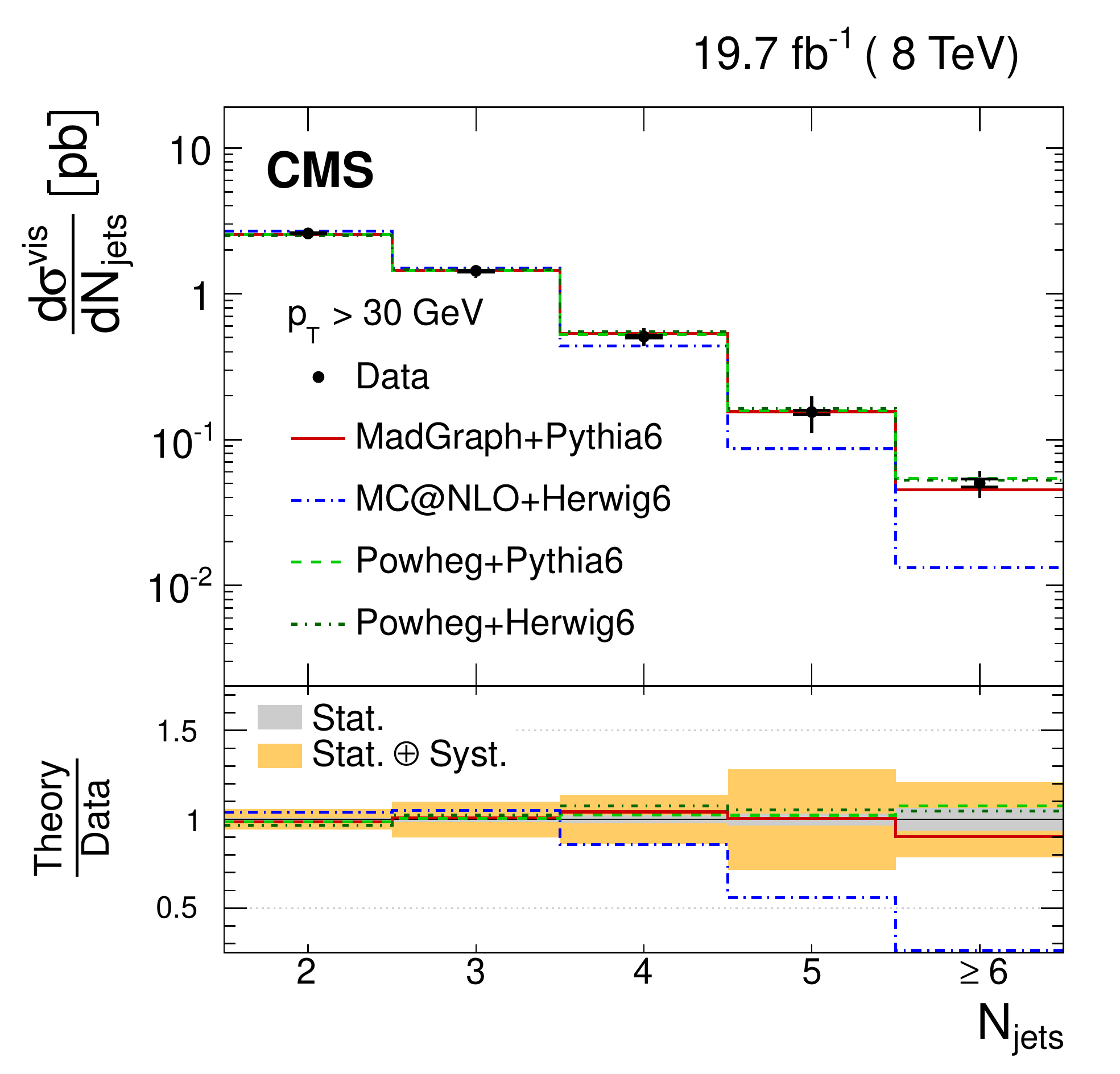}}
\hspace*{1cm}
\subfigure[]{\includegraphics[scale=0.3]{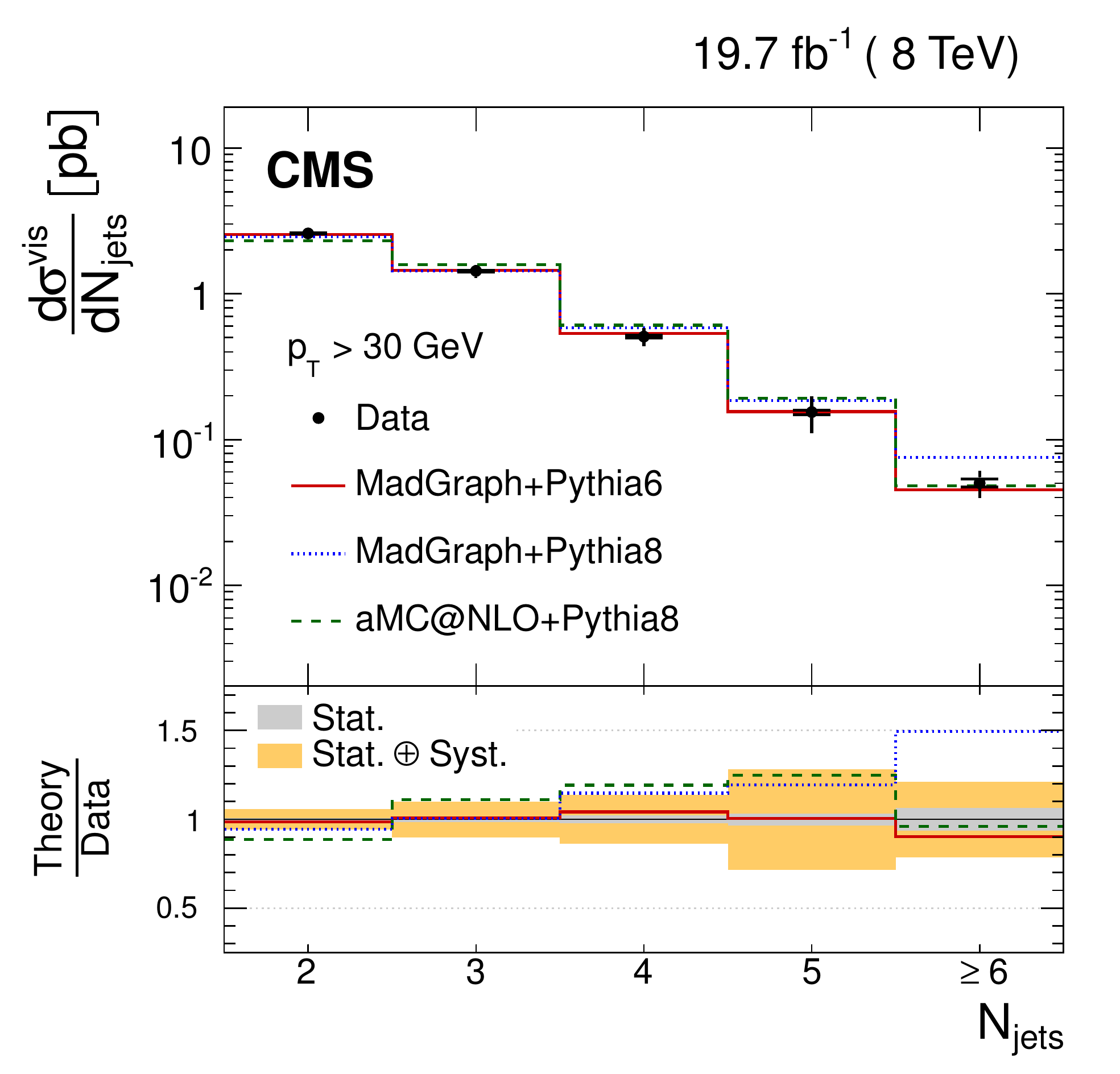}}
\caption{Absolute differential ${\rm t\bar{t}}$ cross sections as a function of jet multiplicity for jets with $p_{\rm T}>30$ GeV. The data are compared with predictions from {\sc MadGraph} + {\sc pythia 6}, {\sc mc@nlo} + {\sc herwig 6}, and {\sc powheg} + {\sc pythia 6} and {\sc herwig 6} (a); and with {\sc MadGraph} +  {\sc pythia 6} and  {\sc pythia 8}, and  {\sc mg5$\_$}a{\sc mc@nlo} +  {\sc pythia 8} (b).}
\label{fig:12_041_njet}
\end{figure}

In order to improve the QCD calculations, it is important to study ${\rm t\bar{t}}$ production in association with jets. Half of the ${\rm t\bar{t}}$ events have extra jets, which are expected to come mainly from ISR. Studying the properties of the jets provide tests of QCD calculations. In Fig.~\ref{fig:12_041_njet} the absolute differential  ${\rm t\bar{t}}$ cross section, measured at $\sqrt{s} = 8$ TeV as a function of jet multiplicity is shown and it is compared to the predictions of {\sc mc@nlo} interfaced with {\sc herwig 6}, which fails at high multiplicities, and {\sc MadGraph} interfaced with {\sc pythia 6} and  {\sc pythia 8}, the latter predicting higher jet multiplicities than the measurement.

\begin{figure}[htb!]
\centering
\subfigure[]{\includegraphics[scale=0.3]{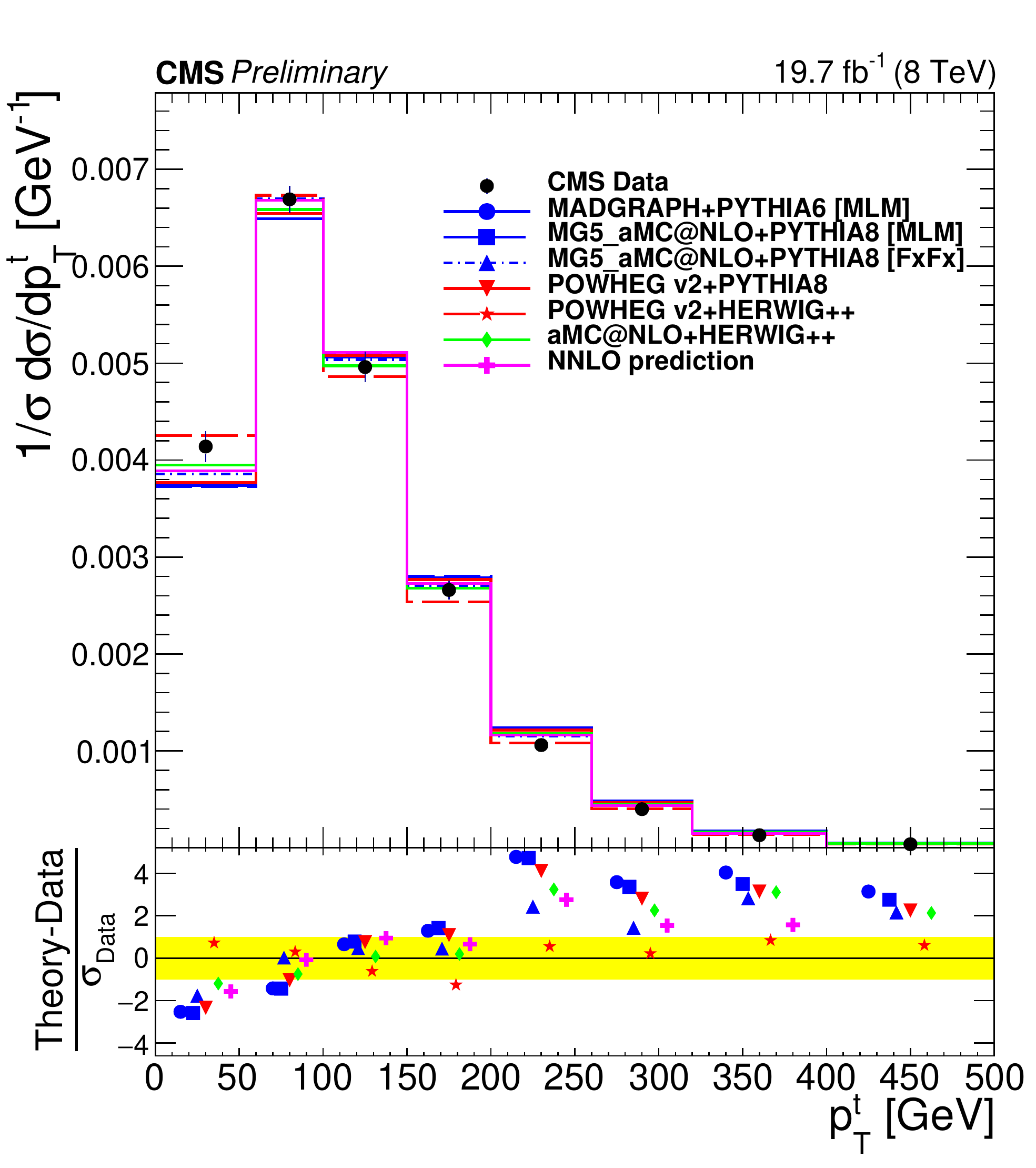}}
\hspace*{1cm}
\subfigure[]{\includegraphics[scale=0.3]{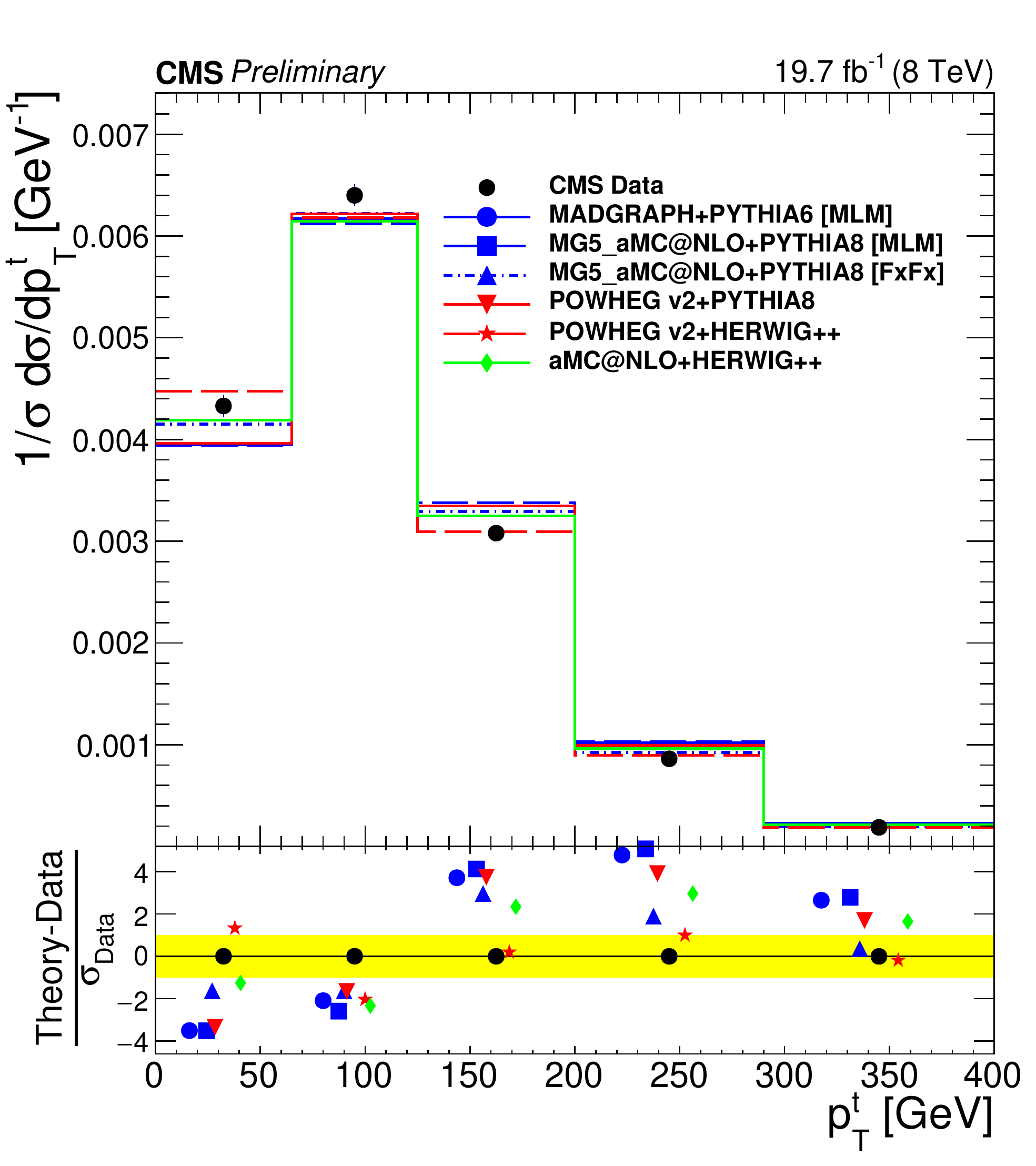}}
\caption{Normalised ${\rm t\bar{t}}$ cross section in bins of $p_{\rm T}^{\rm t}$ in data and MC at the parton-level for lepton + jets (a) and dilepton (b) channels. The yellow band indicates the 1$\sigma$ difference of data from theory predictions. In the lepton+jets channel, the NNLO predictions are also shown.}
\label{fig:15_011_ptt}
\end{figure}

\begin{figure}[htb!]
\centering
\subfigure[]{\includegraphics[scale=0.500]{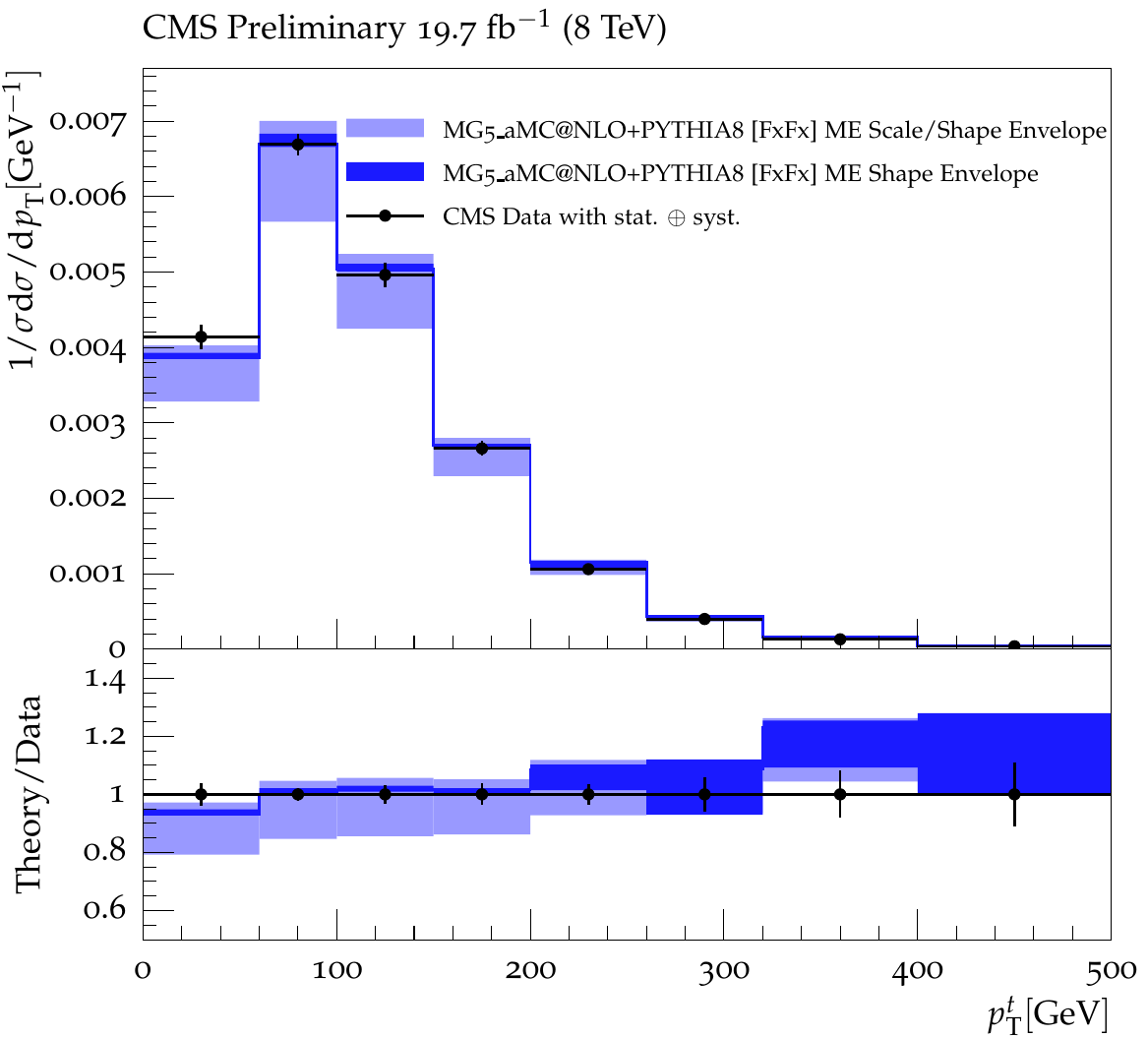}}
\hspace*{1cm}
\subfigure[]{\includegraphics[scale=0.500]{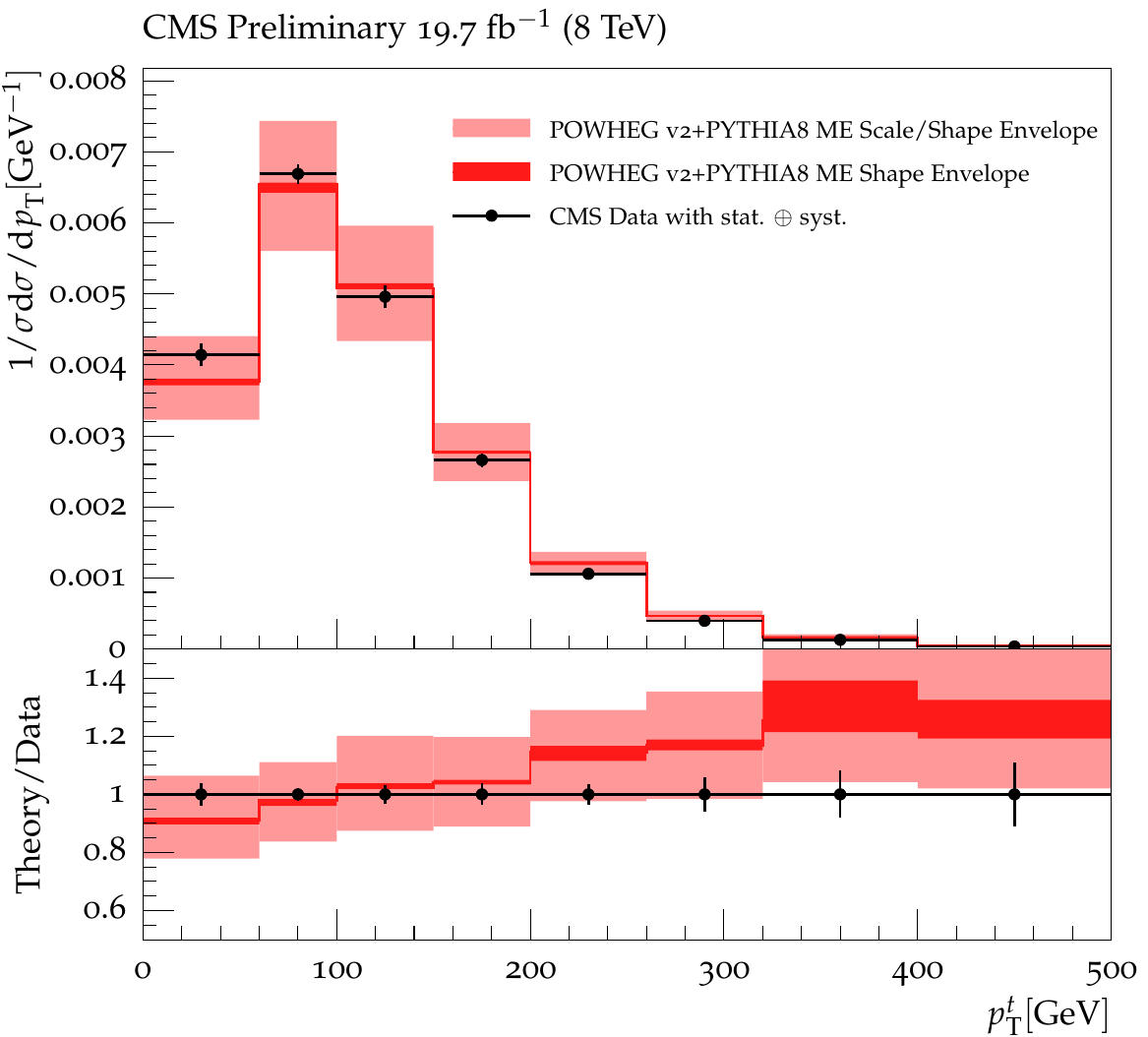}}
\caption{Normalised ${\rm t\bar{t}}$ cross section in bins of $p_{\rm T}^{\rm t}$ in data and {\sc mg5$\_$}a{\sc mc@nlo}+ {\sc pythia 8} [FxFx] sample (a) and {\sc powheg} v2 + {\sc pythia 8} sample (b) at the parton-level for lepton + jets channel. The data points are shown with total error bars and the envelope of different factorization and renormalization assumptions in the matrix elements with a band for both scale+shape and only shape variations.}
\label{fig:15_011_ptt_sc}
\end{figure}

Furthermore, data distributions at 8 TeV are compared with Run 2 generators and parton shower codes, {\sc mg5$\_$}a{\sc mc@nlo} and {\sc powheg} v2 interfaced with {\sc pythia 8} and {\sc herwig++}, using FxFx and MLM matching schemes as shown in Fig.~\ref{fig:15_011_ptt} for the top quark $p_{\rm T}$ at parton level for lepton + jets and dilepton channels. The comparisons include NNLO predictions in lepton + jets channel. In both channels, the best description is obtained by {\sc powheg} +{\sc herwig++} predictions, where the other MC setups show discrepancies.  {\sc mg5$\_$}a{\sc mc@nlo}+ {\sc pythia 8} FxFx sample shows good agreement in the dilepton channel. The impact of the choice of the QCD scale  on the modelling of the top quark $p_{\rm T}$ is shown in Fig.~\ref{fig:15_011_ptt_sc}. The main effect is expected to be observed in the overall normalisation, while the shapes are hardly affected, persisting overall, the same data to MC trend observed for the nominal choice. 

The top $p_{\rm T}$ is mesured at 13 TeV and compared with predictions beyond NLO, as shown in Fig.~\ref{fig:16_011_ptt}. It is observed that beyond NLO predictions seem to give a better description of the top $p_{\rm T}$, consistent with the results obtained in Run I.

\begin{figure}[htb!]
\centering
\subfigure[]{\includegraphics[scale=0.34]{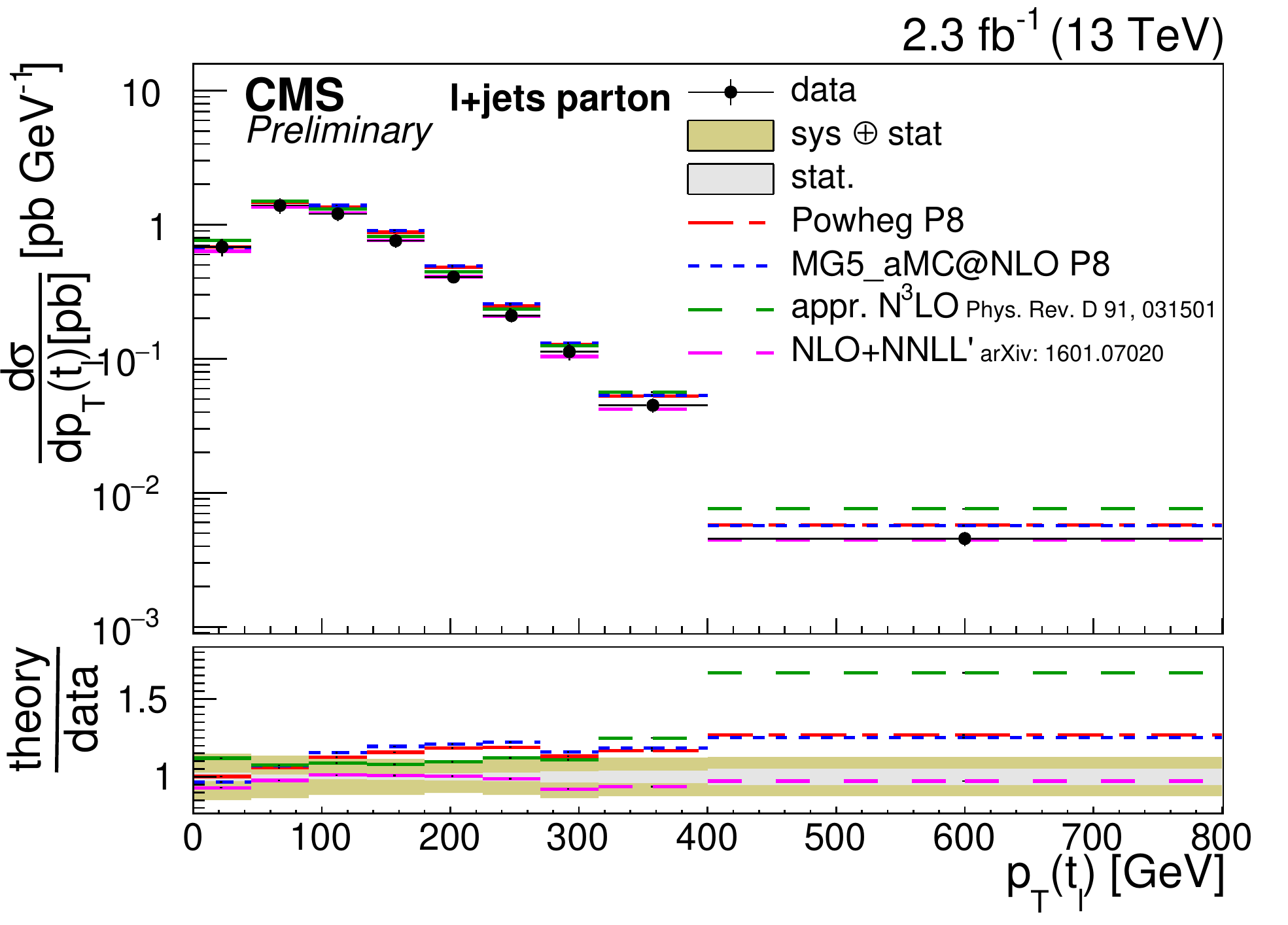}}
\hspace*{1cm}
\subfigure[]{\includegraphics[scale=0.270]{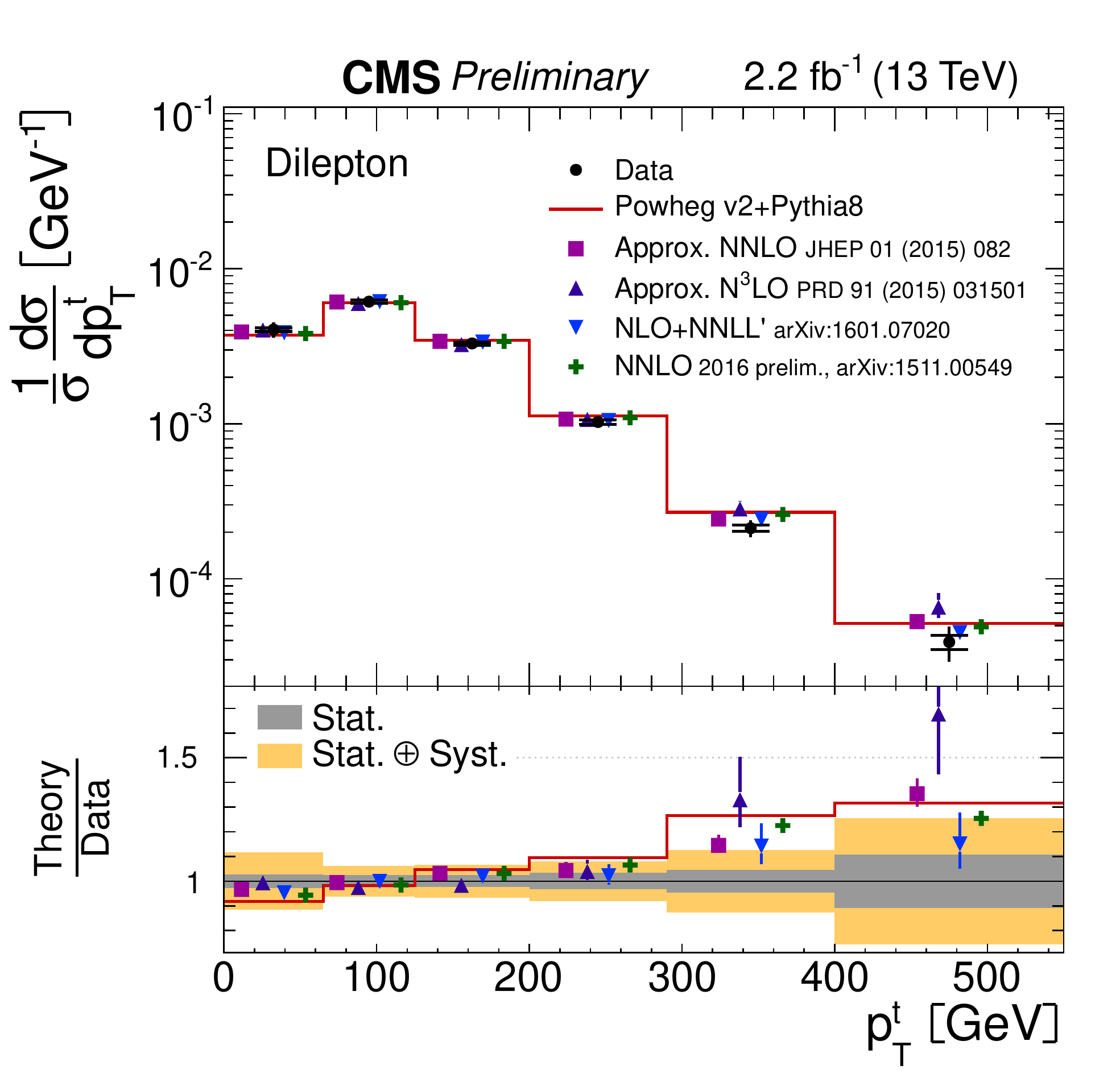}}
\caption{Normalised differential ${\rm t\bar{t}}$ production cross section as a function of $p_{\rm T}^{\rm t}$ for lepton + jets (a) and dilepton (b) channels. \deleted{The inner (outer) error bars indicate the statistical (combined statistical and systematic) uncertainty.}The data are compared to predictions from\deleted{ {\sc powheg} v2  + {\sc pythia 8}, {\sc mg5$\_$}a{\sc mc@nlo} + {\sc pythia 8} [FxFx], {\sc mg5$\_$}a{\sc mc@nlo} + {\sc pythia 8}  [MLM], and {\sc powheg} v2 + Herwig++ (a), and} beyond-NLO QCD calculations.}
\label{fig:16_011_ptt}
\end{figure}

\section{Study of the Underlying Event with ${\rm t\bar{t}}$}
\label{sec:uev}
\begin{figure}[htb!]
\centering
\subfigure[]{\includegraphics[scale=0.22]{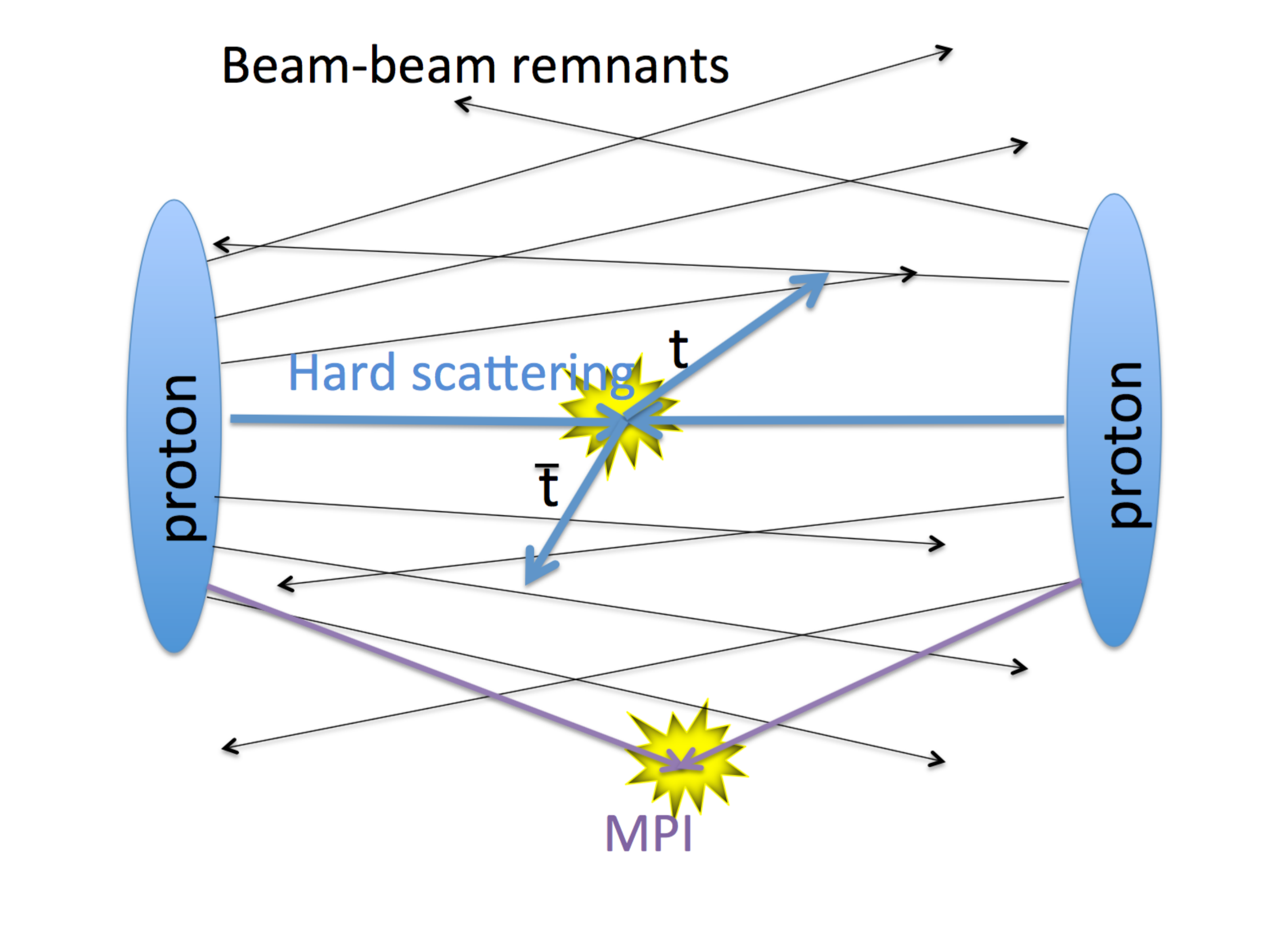}}
\hspace*{1cm}
\subfigure[]{\includegraphics[scale=0.12]{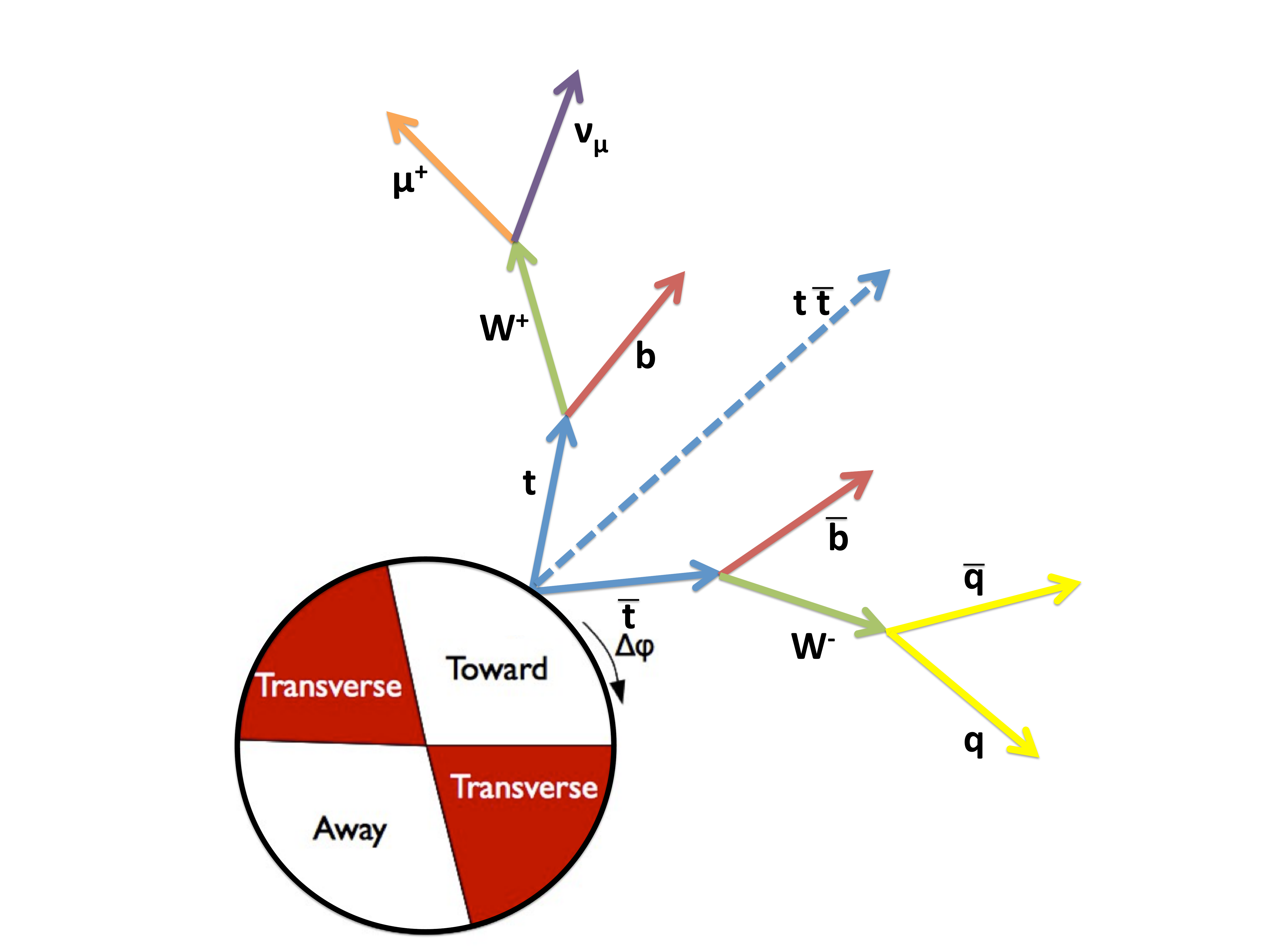}}
\caption{Schematic description of an example proton-proton collision with hard scattering and UE shown (a); and view of the UE regions defined with respect to the azimuthal angle difference between the charged particle candidate and the axis of the ${\rm t\bar{t}}$ system with a schematic diagram of ${\rm t\bar{t}}$ decay chain (b).} 
\label{fig:15_017_uevdesc}
\end{figure}

One important ingredient in the MC simulations is the underlying event (UE), which mainly consists of the beam-beam remnants and the multiple parton interactions (MPI) that accompany the hard scattering, as shown if Fig.~\ref{fig:15_017_uevdesc}a. The UE can not be described by the perturbative QCD calculations and rather relies on phenomenological models whose parameters can be "tuned". 

\begin{figure}[htb!]
\centering
\subfigure[]{\includegraphics[scale=0.34]{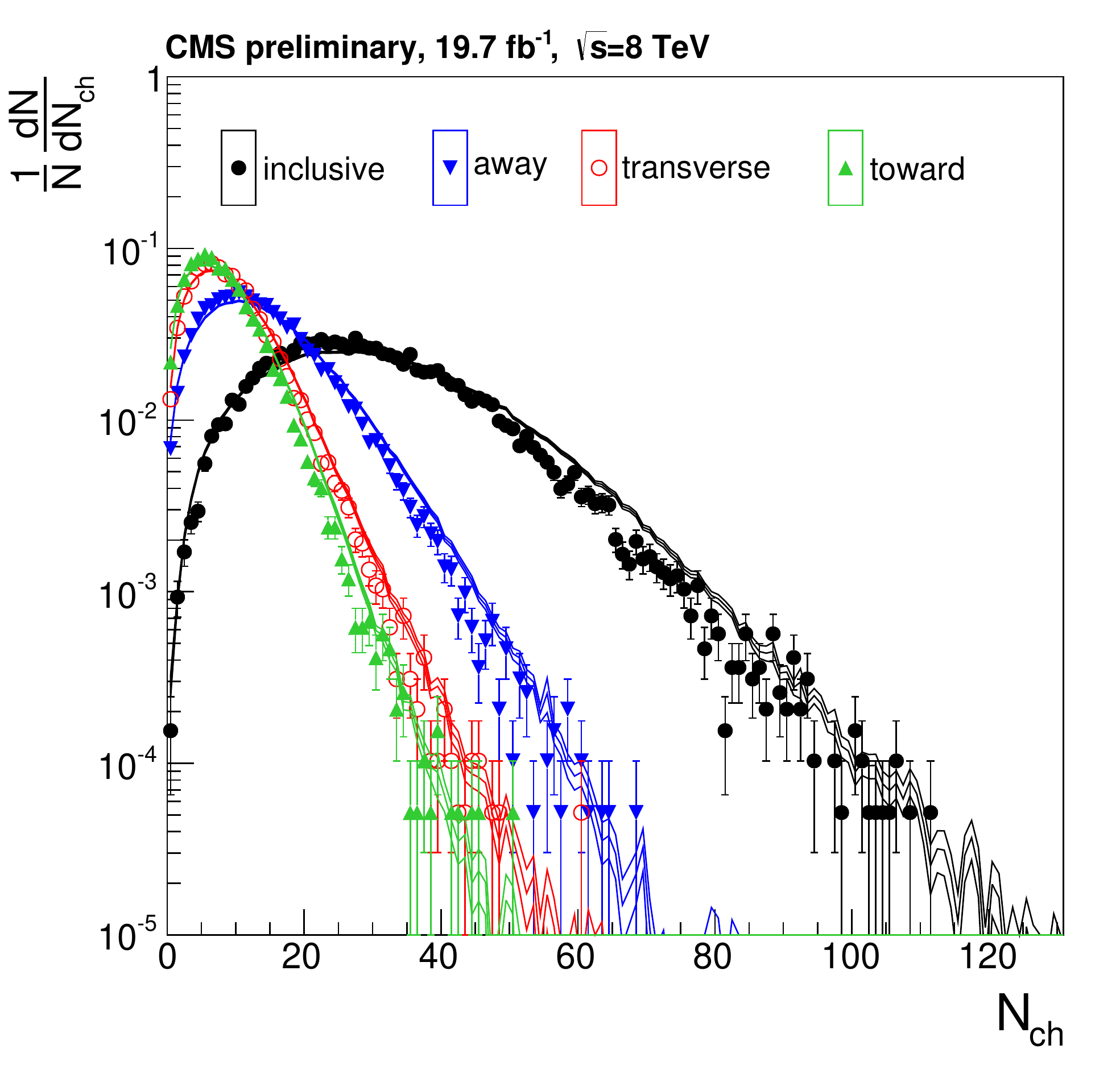}}
\hspace*{1cm}
\subfigure[]{\includegraphics[scale=0.295]{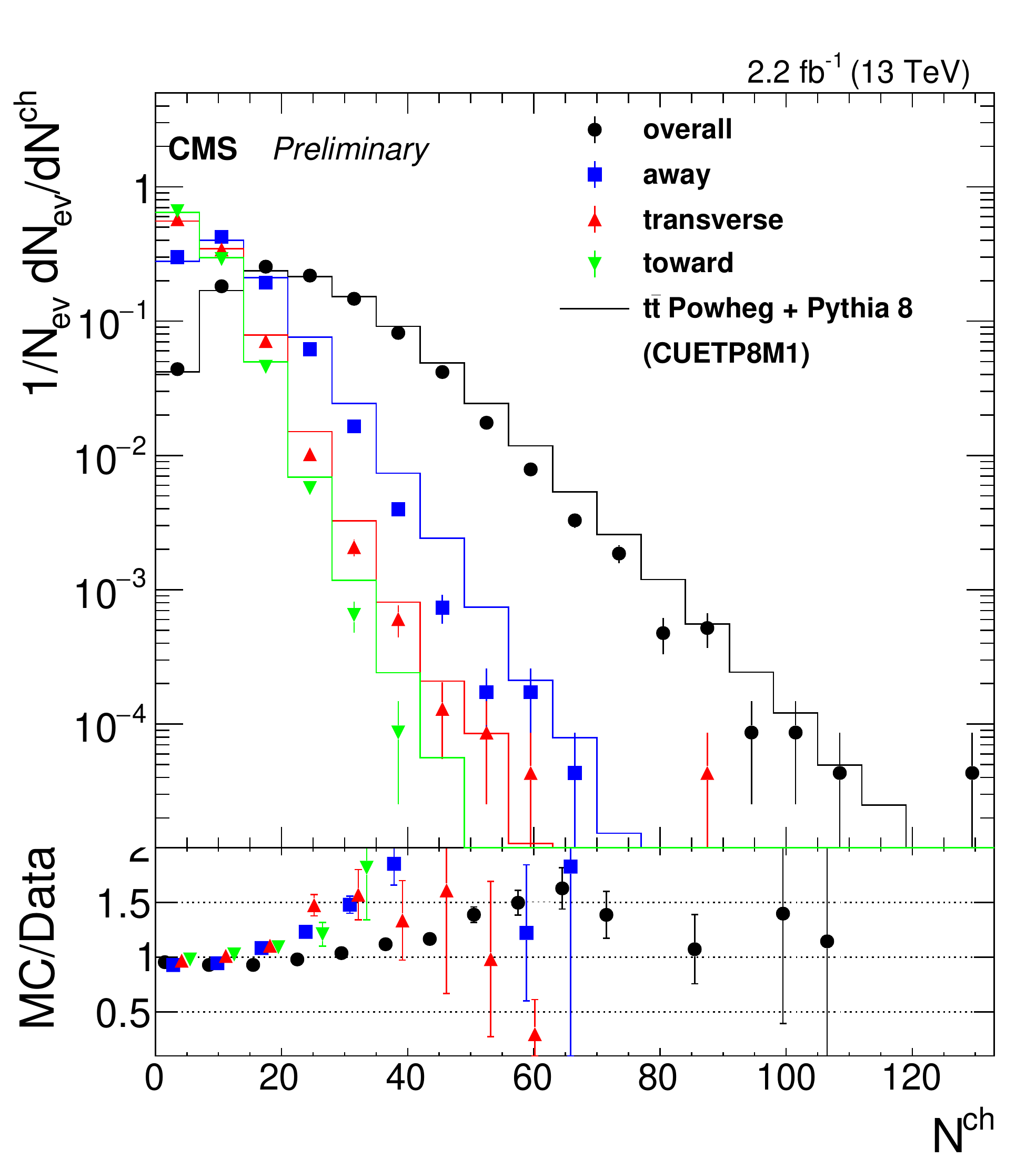}}
\hspace*{1cm}

\caption{The charged PF candidate multiplicity distributions for the away, transverse and toward regions as well as for the overall sample for dilepton channel at 8 TeV (a) and lepton+jets channel at 13 TeV. MC distributions are obtained with the nominal ${\rm Q^2}$ scale.}
\label{fig:15_017_chmult}
\end{figure}

By measuring the UE with  ${\rm t\bar{t}}$, the universality of the tunes at the scale of the ${\rm t\bar{t}}$ process can be verified. Using topological structure of hard hadron-hadron collisions, the UE activity in ${\rm t\bar{t}}$ events is studied at 8 TeV in dilepton and at 13 TeV in lepton+jets final state. The ${\rm t\bar{t}}$  system  is used as the leading  object to define the measurement regions as shown in Fig.~\ref{fig:15_017_uevdesc}b; namely toward, transverse, and away; where the transverse is the most sensitive region to UE activity, and away region contains contents of hard jets accompanying ${\rm t\bar{t}}$. Distributions of number of charged particles are shown for the event regions as well as inclusively for all regions, in Fig.~\ref{fig:15_017_chmult}.  Higher charged particle multiplicity is predicted with the default tunes by MC in both 8 and 13 TeV. It is also observed that  a better description of the data is obtained by MC sample generated with higher parton-shower scale. 

\section{Testing the modelling of b-quark fragmentation}
\label{sec:tmass}

The CMS collaboration has measured the top mass using alternative techniques, using only the kinematic properties of its charged decay products\deleted{. The measured value is $173.68\pm 0.20$ (stat)$^{+1.58}_{-0.97}$(syst) GeV} which minimises the sensitivity to experimental systematic uncertainties. One of the remaining main uncertainties is coming from b-fragmentation modelling, i.e parton to hadron momentum transfer in hadronisation of b quarks. To improve this, previous LEP measurements of b-fragmentation have been used to further tune Z2* for the ${\rm r_b}$ parameter (noted as Z2* LEP ${\rm r_b}$ tune), and used as the nominal b-fragmentation shape. In Fig.~\ref{fig:16_011_ptt} the impact of this choice on the extracted mass value is shown. It is observed that for 1$\%$ change in the average momentum transfer reflects to 0.61 GeV change in the extracted mass value [8]. Further studies comparing the fraction of momentum carried by charmed mesons and secondary vertices reconstructed in b jets in  ${\rm t\bar{t}}$ events, to different fragmentation models, can be found in the same reference.

\begin{figure}[htb!]
\centering
\includegraphics[scale=0.340]{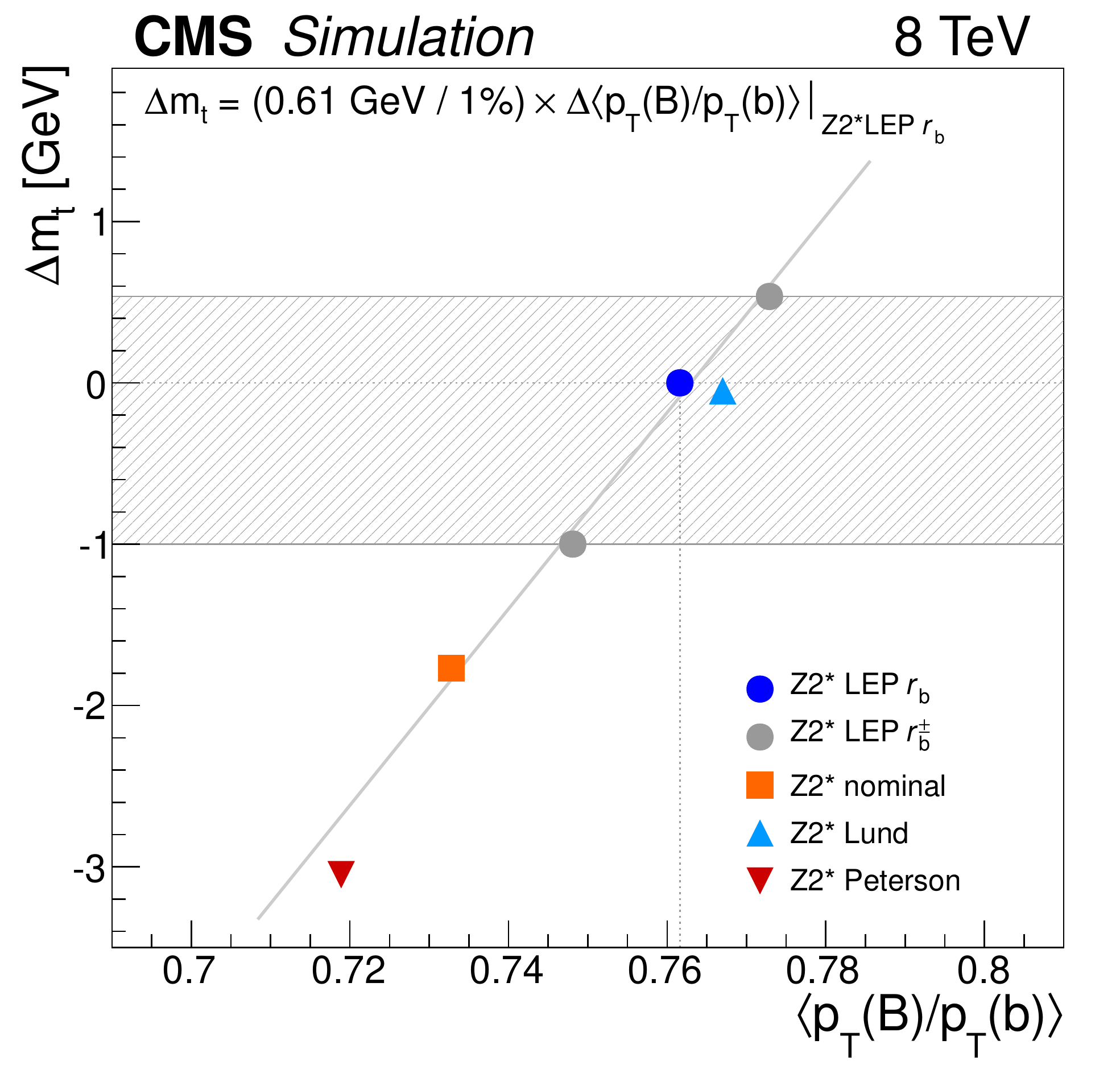}
\caption{Impact of the average b quark fragmentation\deleted{, $<p_{\rm T}{\rm (B)}/<p_{\rm T}{\rm (b)}>$,} on the extracted $m_{\rm t}$ value, for various fragmentation shapes. The horizontal band represents the contribution of the b quark fragmentation model to the systematic uncertainty in the measurement of the top quark mass estimated from the variations of the Z2* LEP ${\rm r_b}$ tune. \deleted{A linear fit to the effects on the different variations (the line in the figure) suggests a relative change in the measured top quark mass of 0.61 GeV for each percent change in average momentum transfer.}}
\label{fig:16_011_ptt}
\end{figure}

\section{Conclusion}

Various CMS measurements are compared to state-of-the-art ME MC generators and parton shower codes. It is seen that the NLO predictions do not always improve the description of data compared to the LO ones. It is also seen that the differences between NLO central value and the measurements are covered by ME scale uncertainty, and better top ${\rm p_{T}}$ description with higher order calculations beyond NLO are observed. Measurements of the UE with ${\rm t\bar{t}}$ shows a preference towards a higher QCD scale choice in simulations, and the UE tunes seem to be universal. The top mass is measured using new technique where the experimental systematics effects are supressed. The dominant remaining uncertainty coming from b-quark fragmentation is expected to be improved by tuning at the LHC.


\end{document}